\documentclass[apjl, iop]{emulateapj}
\usepackage{apjfonts}

\shorttitle{SNR RX~J0852.0$-$4622 with the \emph{Fermi} LAT}
\shortauthors{Tanaka et al.}

\begin{document}

\title{Gamma-Ray Observations of the Supernova Remnant RX~J0852.0$-$4622 with the \emph{Fermi} LAT}


\author{
T.~Tanaka\altaffilmark{1,2}, 
A.~Allafort\altaffilmark{1}, 
J.~Ballet\altaffilmark{3}, 
S.~Funk\altaffilmark{1}, 
F.~Giordano\altaffilmark{4,5}, 
J.~Hewitt\altaffilmark{6}, 
M.~Lemoine-Goumard\altaffilmark{7,8}, 
H.~Tajima\altaffilmark{1,9}, 
O.~Tibolla\altaffilmark{10}, 
Y.~Uchiyama\altaffilmark{1}
}
\altaffiltext{1}{W. W. Hansen Experimental Physics Laboratory, Kavli Institute for Particle Astrophysics and Cosmology, Department of Physics and SLAC National Accelerator Laboratory, Stanford University, Stanford, CA 94305, USA}
\altaffiltext{2}{email: Taka.Tanaka@stanford.edu}
\altaffiltext{3}{Laboratoire AIM, CEA-IRFU/CNRS/Universit\'e Paris Diderot, Service d'Astrophysique, CEA Saclay, 91191 Gif sur Yvette, France}
\altaffiltext{4}{Dipartimento di Fisica ``M. Merlin" dell'Universit\`a e del Politecnico di Bari, I-70126 Bari, Italy}
\altaffiltext{5}{Istituto Nazionale di Fisica Nucleare, Sezione di Bari, 70126 Bari, Italy}
\altaffiltext{6}{NASA Goddard Space Flight Center, Greenbelt, MD 20771, USA}
\altaffiltext{7}{Universit\'e Bordeaux 1, CNRS/IN2p3, Centre d'\'Etudes Nucl\'eaires de Bordeaux Gradignan, 33175 Gradignan, France}
\altaffiltext{8}{Funded by contract ERC-StG-259391 from the European Community}
\altaffiltext{9}{Solar-Terrestrial Environment Laboratory, Nagoya University, Nagoya 464-8601, Japan}
\altaffiltext{10}{Institut f\"ur Theoretische Physik and Astrophysik, Universit\"at W\"urzburg, D-97074 W\"urzburg, Germany}

\begin{abstract}
We report on gamma-ray observations of the supernova remnant (SNR) RX~J0852.0$-$4622 with the Large Area Telescope (LAT) 
aboard the \emph{Fermi Gamma-ray Space Telescope}. In the \emph{Fermi} LAT data, we find a spatially extended 
source at the location of the SNR. The extension is consistent with the SNR size seen in other wavelengths such as 
X-rays and TeV gamma rays, leading to the identification of the gamma-ray source with the SNR. The spectrum 
is well described as a power law with a photon index of  $\Gamma = 1.85 \pm 0.06~{\rm (stat)} ^{+0.18}_{-0.19}~{\rm (sys)}$, which 
smoothly connects to the H.E.S.S. spectrum in the TeV energy band. We discuss the gamma-ray emission mechanism 
based on multiwavelength data.  The broadband data can be fit well by a model in which the gamma rays are of hadronic origin. 
We also consider a scenario with inverse Compton scattering of electrons as the emission mechanism of the gamma rays.  
Although the leptonic model predicts a harder spectrum in the \emph{Fermi} LAT energy range, the model can fit the data 
considering the statistical and systematic errors.  
\end{abstract}


\section{Introduction}
Shell-type supernova remnants (SNRs) are prime candidates for acceleration sites of Galactic cosmic rays \citep{blandford78, malkov01}. 
The energy density of cosmic rays ($\sim 1~{\rm eV}~{\rm cm}^{-3}$) can be explained if $\sim 10$\% of 
the kinetic energy released by supernova explosions is transferred to accelerated particles. 
Also, diffusive shock acceleration theory, which is expected to be working at SNR shocks, can naturally 
explain the power-law type spectrum of cosmic rays. 
Observationally, nonthermal X-rays and TeV gamma rays have been detected from SNRs, particularly from 
relatively young ones such as Cassiopeia A \citep[e.g.][]{hughes00, vink03, albert07}, 
RX~J1713.7$-$3946 \citep[e.g.][]{tanaka08, acero09, hess_1713}, and 
RX~J0852.0$-$4622 \citep[e.g.][]{slane01, iyudin05, pannuti10, hess07}, 
providing us with evidence that particles are indeed accelerated at least up to $\sim 100$~TeV. 

The SNR RX~J0852.0$-$4622 (also known as G266.2$-$1.2 or Vela Jr.) is one of the SNRs from which both 
nonthermal X-rays and TeV gamma rays are detected. 
The SNR was discovered in X-rays by \emph{ROSAT} in the southeastern corner
of the Vela SNR \citep{aschenbach98}. Its X-ray spectrum was later confirmed to be of nonthermal origin by \emph{ASCA} observations \citep{slane01}. 
The X-ray emission has an almost circular morphology with a diameter of $\sim 2^\circ$ with rim brightening 
especially in the north, northwestern, and western parts. 
At TeV energies, the CANGAROO \citep{katagiri05,enomoto06} and H.E.S.S. \citep{hess05,hess07} collaborations reported the detection. 
Similar to the well-studied SNR RX~J1713.7$-$3946, the H.E.S.S. telescope spatially resolved the emission from 
RX~J0852.0$-$4622 \citep{hess07}. 

The most recent and reliable estimates of age and distance of the SNR 
come from the measurement of expansion of the shell by \cite{katsuda08} 
using \emph{XMM-Newton} data taken over a span of 6.5~yr. 
Their estimates of the age and distance are 1700--4300~yr and 750~pc, respectively. 
At this distance, the angular size of the SNR ($\sim 1^{\circ}$ in radius) translates to a radius of  $\sim 13$~pc. 
It is widely believed that RX~J0852.0$-$4622 is a remnant of a core-collapse (type Ib/Ic/II) supernova explosion 
since studies of an X-ray point source, CXOU~J085201.4$-$461753, located close to the SNR center 
suggest that the source is a neutron star \citep[e.g.][]{pavlov01,kargaltsev02}. 
However, \cite{reynoso06} pointed out the X-ray source is more likely related to a planetary nebula 
they found at the location in radio data. 

 The broadband spectrum of the nonthermal emission can be explained either by a 
leptonic scenario where the gamma rays are produced by inverse Compton scattering of high-energy electrons,  
or by a hadronic scenario where $\pi^0$ decays are responsible for the gamma rays \citep{hess07}. 
The Large Area Telescope (LAT) aboard the \emph{Fermi Gamma-ray Space Telescope} \citep{atwood09}
can provide important information on GeV gamma-ray emission which is crucial to constrain the possible 
emission models. 
In this Letter, we report on results of gamma-ray observations of the SNR RX~J0852.0$-$4622 
by the \emph{Fermi} LAT.

\section{Observations and Data Reduction}
The \emph{Fermi Gamma-ray Space Telescope}, launched on 2008 June 11, started 
continuous survey-mode observations on 2008 August 4. 
The LAT, with its wide field-of-view (FoV) of 2.4~sr, 
observes the whole sky in every $\sim 3$~hr in this mode. 
Besides the wide FoV, the LAT has better detector performance than 
its predecessors, large effective area of $\sim 8000~{\rm cm}^2$ (on axis at 1~GeV) 
and point spread function better than $1.0^\circ$ (for 68\% containment) at 1~GeV. 
A detailed description of the instrument can be found in \cite{atwood09}. 

Here we analyzed {\em Fermi} LAT data taken for 30 months, from 2008 August 4 to 2011 February 8. 
The data analysis was performed using the standard analysis software package, the LAT Science Tools v9r21p0. 
We selected ``Diffuse class'' events, which are those reconstructed with high 
probability of being photons and are recommended for standard point-source analyses.  
In the analysis, the instrumental response functions (IRFs) ``Pass6 v11'' were used. 
To reduce the background from the Earth limb \citep{albedo}, 
events with Earth zenith angle greater than $100^\circ$ were eliminated.

\section{Analysis and Results}
We performed a morphological and spectral analysis of the region that contains SNR RX~J0852.0$-$4622. 
All the observational results except for the counts map shown in Figure~\ref{fig:map} (a) and (b) are based on binned maximum likelihood analysis. 
The emission model for the maximum likelihood analysis includes SNR RX~J0852.0$-$4622, Vela X, 
point sources listed in the First \emph{Fermi} LAT (1FGL) catalog \citep{1fgl}, the Galactic diffuse background, and 
the isotropic background component which accounts for the extragalactic diffuse emission and residual detector background. 
One of the 1FGL sources, 1FGL~J0854.0$-$4632, is located inside the SNR RX~J0852.0$-$4622, and was excluded from the model. 
The spatial template for Vela X is assumed to be a uniform disk as reported by \cite{velax}. 
The Galactic diffuse and isotropic background models we adopted are {\it gll\_iem\_v02\_P6\_V11\_DIFFUSE.fit} and 
{\it isotropic\_iem\_v02\_P6\_V11\_DIFFUSE.txt}\footnote{http://fermi.gsfc.nasa.gov/ssc/data/access/lat/BackgroundModels.html}, respectively. 
The analysis was performed on events within a region-of-interest (ROI) of a $14^\circ \times 14^\circ$ square centered 
at  ($\alpha_{\rm J2000}$, $\delta_{\rm J2000})$ = ($133 \fdg 00$, $-46 \fdg 37$).

\subsection{Spatial Distribution and Identification}
\begin{figure*}[bhpt]
\epsscale{0.5}
\plotone{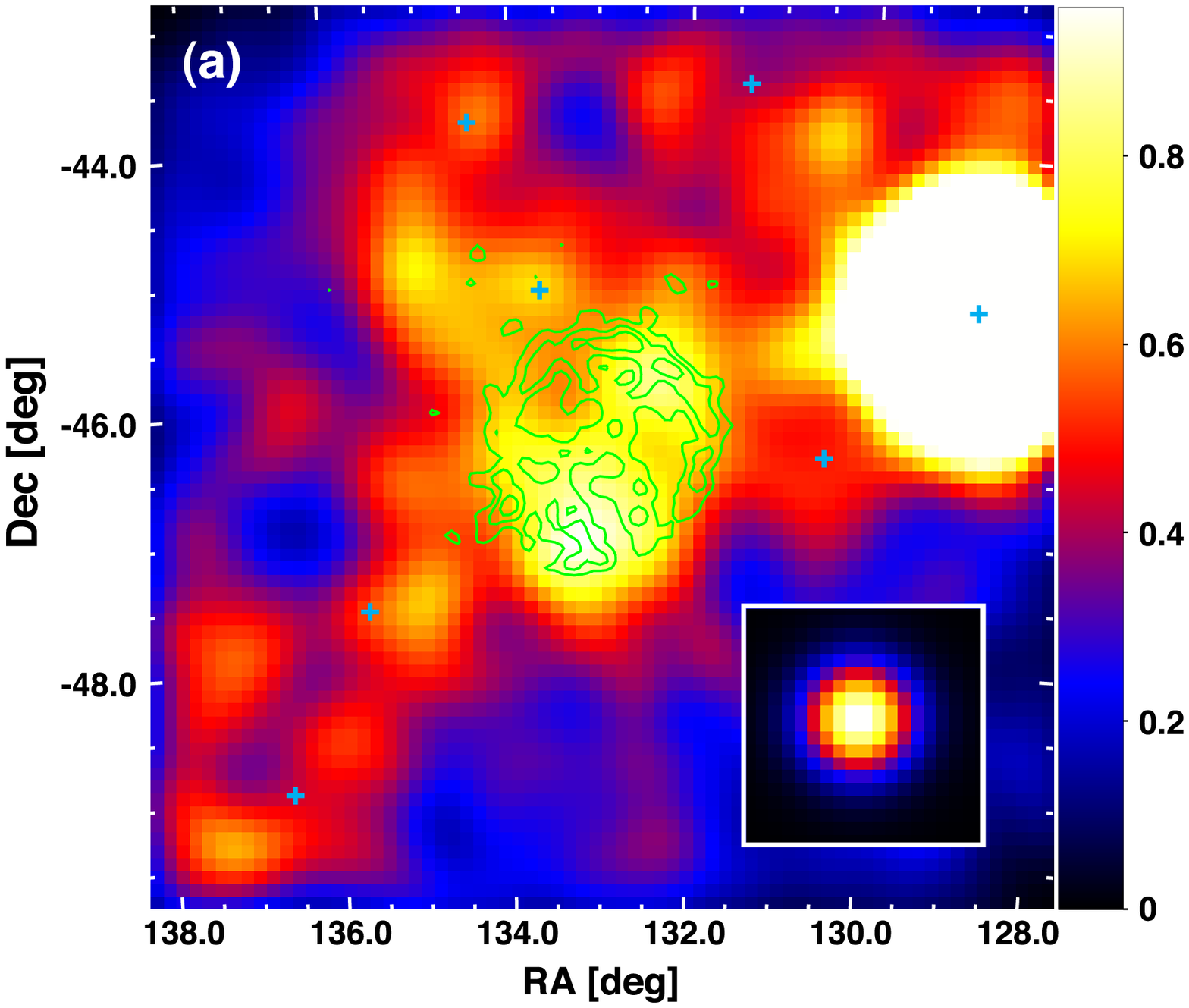}
\plotone{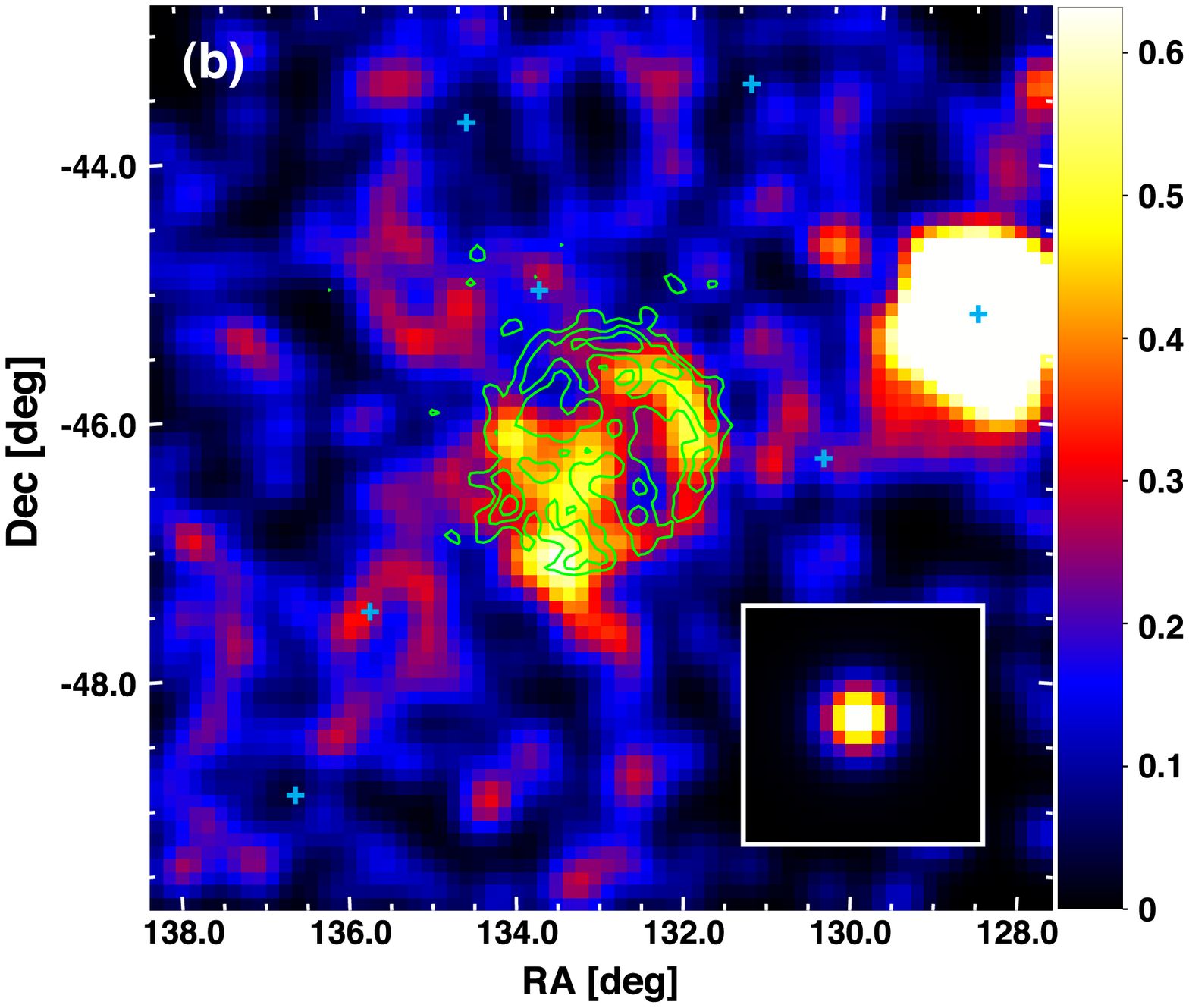}
\plotone{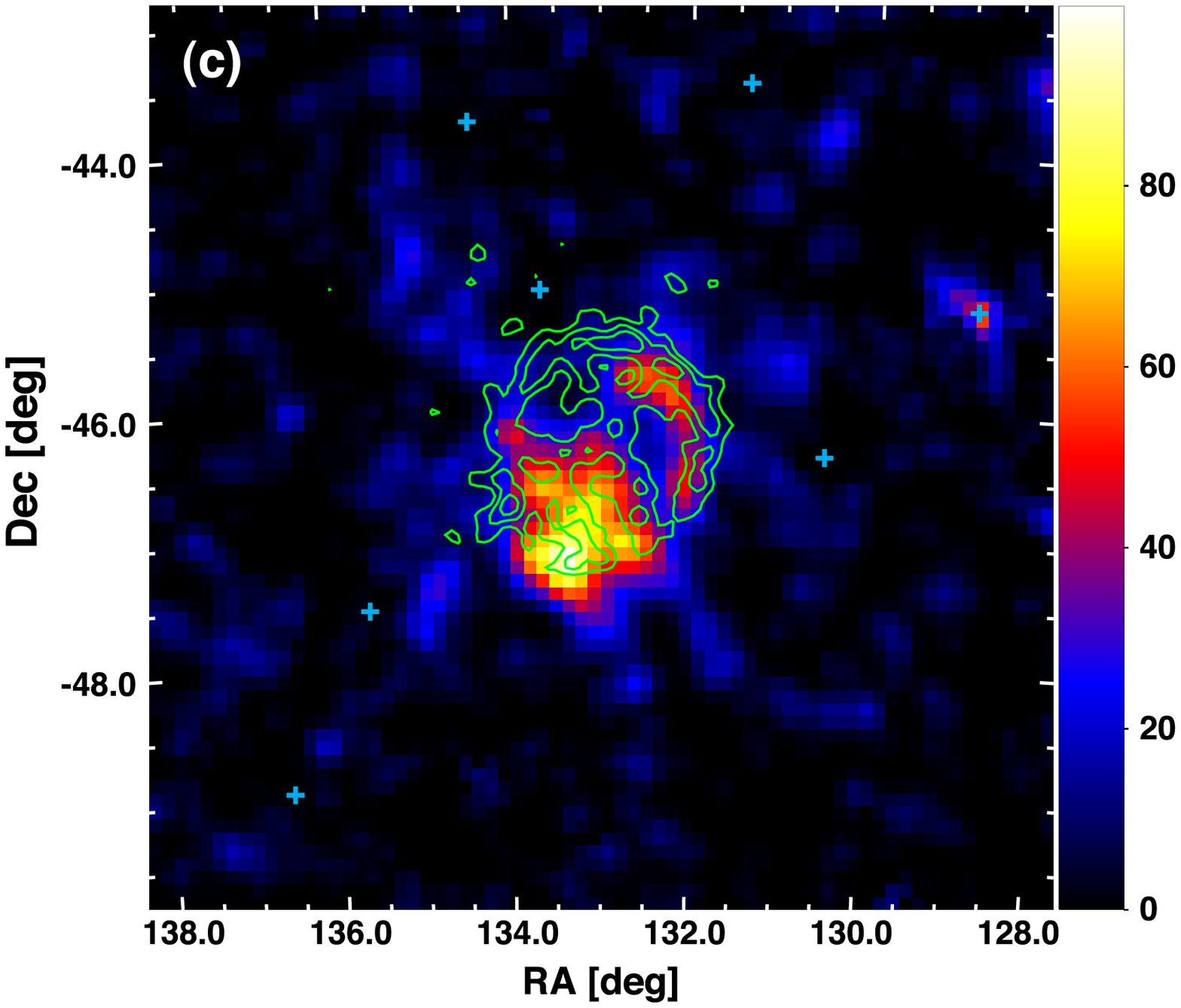}
\caption{\small  (a) Counts map of $> 5$~GeV events. 
The unit of the color scale is counts per pixel with a pixel size of $0^\circ.1 \times 0^\circ.1$. 
The green contours are the H.E.S.S. image \citep{hess07}. 
Locations of 1FGL catalog sources included in the emission model are indicated with the cyan crosses. The brightest source located to the west of the SNR is the Vela pulsar \citep{vela_pulsar}. The map is smoothed with a Gaussian kernel with $\sigma = 0 \fdg 25$. The inset is an image of a simulated point source with the same spectral shape as the SNR. (b) Counts map of $> 10$~GeV events. Gaussian smoothing is applied with $\sigma = 0 \fdg 15$. (c) Map of test statistic for events above 5~GeV. 
\label{fig:map} }
\end{figure*}

In counts maps of \emph{Fermi} LAT data, a gamma-ray source was found at the location of SNR RX~J0852.0$-$4622. 
Figure~\ref{fig:map} (a) shows a \emph{Fermi} LAT counts map of $> 5$~GeV events. 
To take advantage of better angular resolution and weaker Galactic diffuse emission, we 
also present a counts map of $> 10$~GeV events in Figure~\ref{fig:map} (b). 
No  background subtraction is applied to the counts maps. 
Emission can be seen at the location of SNR RX~J0852.0$-$4622. 
Figure~\ref{fig:map} (c) presents the test statistic (TS) map of the source calculated using events above 10~GeV. 
TS is defined as ${\rm TS} =  -2 \ln (L_0/L_{\rm point})$, where $L_0$ and $L_{\rm point}$ are the likelihood of the null hypothesis 
(without a source) and that of a point source hypothesis, respectively \citep{mattox96}.  
To generate the TS map, we put a point source at each position in the sky and obtained TS for each case. 
The TS map indicates significant emission from the location coincident with RX~J0852.0$-$4622 above the backgrounds. 

The spatial distribution of the emission shown in Figure~\ref{fig:map} appears to be extended and 
compatible with the SNR seen in other wavelengths. 
To test the spatial extension and quantify it, we assumed a uniform disk as a spatial template and 
performed a maximum likelihood fit to $> 5$~GeV events with {\it pointlike} \citep{extended_cat}. 
The emission was found to be extended significantly with ${\rm TS}_{\rm ext}  \equiv -2 \ln (L_{\rm point}/L_{\rm disk}) = 144$, where 
$L_{\rm disk}$ is the likelihood for the case in which a disk is adopted as a spatial template for the source. 
The best fit radius is $1 \fdg 12^{+0 \fdg 07}_{-0 \fdg 06}$ centered at ($\alpha_{\rm J2000}$, $\delta_{\rm J2000}$) = ($133 \fdg 2 \pm 0 \fdg 1$, $-46 \fdg 52 \pm 0 \fdg 08 $). 
The TS for the source detection is 228 with the best-fit disk template. 
The location and size are consistent with those of SNR RX~J0852.0$-$4622 seen in radio, X-rays, and TeV gamma rays, 
and, therefore, support the identification of the emission with the SNR. 
We also tried fitting the data using the H.E.S.S. image \citep{hess07} as a spatial template. 
In this case, we obtained ${\rm TS} = 221$, which is only slightly different from that obtained for the disk shape. 
The current statistics do not allow us to study further the morphology of the GeV emission such as 
spatial structures inside the SNR. 
In what follows, we use the H.E.S.S. image as the default spatial 
template to study the spectral properties of the source. 

\subsection{Spectrum}
The spectral information of the source was extracted by performing maximum likelihood fits with {\it gtlike}\footnote{http://fermi.gsfc.nasa.gov/ssc/data/analysis/} for the ROI. 
In the energy region below 1~GeV, we found that the systematic errors mainly from the Galactic diffuse model 
become $\gtrsim 80$\% of the flux obtained for RX~J0852.0$-$4622, and thus we only report results obtained above 1~GeV. 
Firstly, we analyzed all the events above 1~GeV together to obtain the spectral parameters 
in the entire band.
We assumed a power-law function as a spectral model for RX~J0852.0$-$4622. 
The 1FGL sources except for the Vela pulsar were also modeled with power laws and their parameters are allowed to vary. 
The spectrum of the Vela pulsar is represented with a power law with an exponential cutoff, whose parameters (normalization, index, and cutoff energy) 
are also treated as free parameters.  
To the spectrum of the Galactic diffuse model, a power-law function was multiplied in order to allow for 
any energy-dependent discrepancy between the model and data. 
The photon index of RX~J0852.0$-$4622 is $\Gamma = 1.85 \pm 0.06~{\rm (stat)} ^{+0.18}_{-0.19}~{\rm (sys)}$, and  
the flux integrated between 1~GeV and 300~GeV is 
$(1.01 \pm 0.09~{\rm (stat)} ^{+0.36}_{-0.31}~{\rm (sys)}) \times 10^{-8}~{\rm ph}~{\rm cm}^{-2}~{\rm s}^{-1}$. 

\begin{figure}[htbp]
\epsscale{1.1}
\plotone{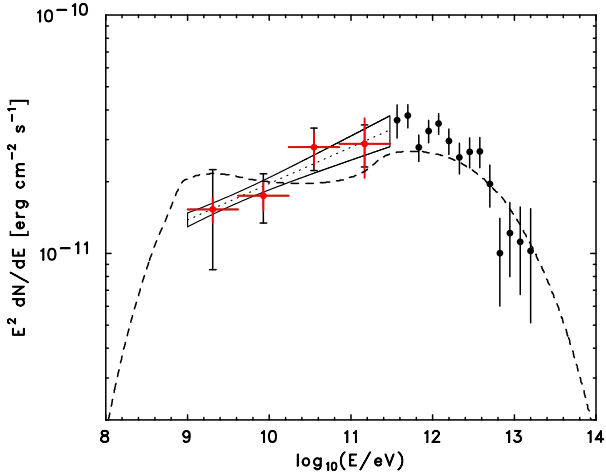}
\caption{\small  \emph{Fermi} LAT spectral energy distribution (SED) in 1--300~GeV with the H.E.S.S. SED by \cite{hess07} plotted together. For the \emph{Fermi} LAT points, the vertical red lines  and the black caps represent statistical and systematic errors, respectively. The dotted line indicates the best-fit power law obtained from the maximum likelihood fit for the entire 1--300~GeV band. The butterfly shape shows the 68\% confidence region. The dashed curve is the  $\pi^0$-decay spectrum by \cite{berezhko09}. 
\label{fig:spec} }
\end{figure}

In order to obtain spectral points at each energy, we divided the data between 1~GeV and 300~GeV 
into four log-equal energy bands, and performed maximum likelihood fits for each of them. 
The parameters for the point sources in the ROI, except for the Vela pulsar, are fixed at the best-fit values obtained in the entire-band fit above. 
The fitting parameters here are the normalization of RX~J0852.0$-$4622, that of the Vela pulsar, and that of the Galactic diffuse emission. 
The spectral points obtained in each band are summarized in Table~\ref{tab:spec_points} and are plotted in Figure~\ref{fig:spec} together 
with the power-law function from the entire band fit. 

The major contributions to the systematic errors quoted above come from uncertainties in the effective area calibration of the \emph{Fermi} LAT 
and imperfect modeling of the Galactic diffuse emission. 
Uncertainties in the effective area are estimated based on observations of the Vela pulsar \citep{vela_pulsar} 
and Earth limb gamma rays \citep{albedo}. 
Here we assumed 5\% at $\log_{10}(E/{\rm MeV}) = 2.75$, 20\% above 
$\log_{10}(E/{\rm MeV}) = 4$, and interpolation of the two values in-between. 
The accuracy of the Galactic diffuse model was evaluated based on the discrepancies between the best-fit model and 
data at each location of the ROI as was done by \cite{w51c}. 
We found that they are $\lesssim 10\%$, and we changed the normalization of the Galactic diffuse model by $\pm 10\%$ from the best-fit values and added 
the differences of the parameters before and after the renormalization to the systematic errors. 

\begin{deluxetable*}{ccc}
\tablecolumns{9}
\tablewidth{0pt}
\tablecaption{\emph{Fermi} LAT Spectral Points\label{tab:spec_points}}
\tablehead{
$E$ [GeV] & $E^2\, dN(E)/dE$ [${\rm erg}~{\rm cm}^{-2}~{\rm s}^{-1} $]  & TS
 }
\startdata
2.04 & $(1.5 \pm 0.2 \pm 0.7) \times 10^{-11}$ & 84.2  \\
8.49 & $(1.7 \pm 0.2 \pm 0.4) \times 10^{-11}$ & 86.1  \\
35.3 & $(2.8 \pm 0.4 \pm 0.6) \times 10^{-11}$ & 102.4 \\
147  & $(2.9 \pm 0.8 \pm 0.6) \times 10^{-11}$  &  40.9 
\enddata
\tablecomments{The first and second errors denote statistical and systematic errors, respectively.}
\end{deluxetable*}

\begin{figure}[htbp]
\epsscale{1.2}
\plotone{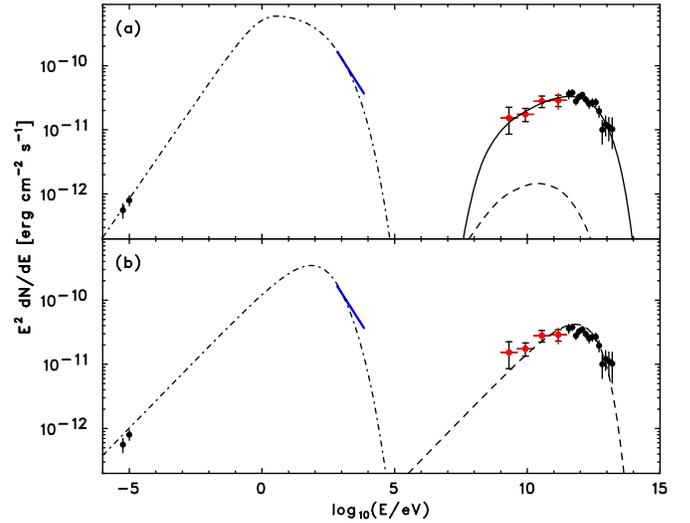}
\caption{\small  Broadband SED of RX~J0852.0$-$4622 with (a) the hadronic model and (b) the leptonic model. The radio data points are integrated fluxes of the SNR determined based on the 64-m Parkes radio telescope data by \cite{duncan00}. The blue line shows the X-ray flux estimated by \cite{hess07}, who analyzed \emph{ASCA} data. The solid, dashed, and dot-dashed curves represent contributions from $\pi^0$ decays, inverse Compton scattering, and synchrotron radiation, respectively. The parameters adopted for the model calculation are summarized in Table~\ref{tab:model_param}.
\label{fig:modeling} }
\end{figure}

\begin{deluxetable*}{ccccccccc}
\tablecolumns{9}
\tablewidth{0pt}
\tablecaption{Parameters for the Models \label{tab:model_param}}
\tablehead{
Model & $s_p$ & $p_{0p}$ & $s_e$ & $p_{0e}$ & $B$ & $n$ & $W_p$ & $W_e$ \\
 &     &  [${\rm TeV}~c^{-1}$] &    &  [${\rm TeV}~c^{-1}]$ & [$\mu{\rm G}]$ & [${\rm cm}^{-3}] $ & [erg] & [erg] 
 }
\startdata
Hadronic & 1.8 & 50 & 1.8 & 10 & 100 & 0.1 & $5.2 \times 10^{50}$ & $3.9 \times 10^{46}$ \\
Leptonic & --- & --- & 2.15 & 25 & 12 & 0.1 & --- &  $1.1 \times 10^{48}$ 
\enddata
\tablecomments{$W_p$ and $W_e$ are total kinetic energy of particles integrated above 10~MeV.}
\end{deluxetable*}

\section{Discussion}
Gamma-ray emission mechanisms for RX~J0852.0$-$4622 
have been actively discussed in the literature since its detection in the TeV band. 
Two types of scenarios have mainly been considered to account for the nonthermal emission from RX~J0852.0$-$4622 \citep[e.g.][]{hess07}. 
One is the so-called hadronic scenario in which gamma rays are dominantly radiated through decays of $\pi^0$ mesons 
produced in collisions between accelerated protons/nuclei and ambient gas. 
The other is the leptonic scenario in which inverse Compton scattering of relativistic electrons with photon fields 
such as the cosmic microwave background (CMB) is the main emission mechanism for gamma rays. 

We calculated gamma-ray emission models for both the hadronic and leptonic scenarios.  
In the calculations, protons and electrons are injected with a constant rate and fixed spectral form of 
\begin{eqnarray}
Q_{e, p} = A_{e, p} \left( \frac{p}{1~{\rm GeV}~c^{-1}} \right)^{-s} \exp \left[-\left( \frac{p}{p_0} \right)\right]. 
\end{eqnarray}
Injected particles suffer cooling via such processes as ionization, synchrotron, or inverse Compton scattering, 
and their spectra are deformed. 
The spectra of radiating particles ($N_{e,p}$) are obtained by solving the kinetic equations for electrons and protons:
\begin{eqnarray}
\frac{\partial N_{e,p}(p, t)}{\partial t} = \frac{\partial}{\partial p} [b_{e,p}(p)\ N_{e,p}(p, t)] + Q_{e,p}(p), 
\end{eqnarray}
where the term $b_{e,p}$ denotes the momentum loss rate. 
The equations are solved for $t = 3000$~yr, and then photon spectra from electrons and protons are computed for each radiation process. 
The distance to the SNR is assumed to be $750$~pc \citep{katsuda08}. 
As the seed photons for inverse Compton scattering, we considered the CMB as well as the interstellar radiation field in infrared 
and optical bands. The spectrum of the interstellar radiation is taken from GALPROP \citep{porter06} at the location of the SNR.

Figure~\ref{fig:modeling} (a) shows the hadronic model together with 
radio, X-ray, and gamma-ray data. 
The parameters adopted for the calculation are summarized in Table~\ref{tab:model_param}. 
The indices of proton and electron spectra are assumed to be the same. 
For the model curves in Figure~\ref{fig:modeling} (a), the proton spectrum has an index 
of $s_p = 1.8$ and a high-energy cutoff at $p_{0p} = 50~{\rm TeV}~c^{-1}$. In this case, the total kinetic energy of protons above 10~MeV is 
$W_p = 5.2 \times 10^{50}~(n/0.1~{\rm cm}^{-3})^{-1}$~erg, where $n$ denotes the hydrogen gas density in the post-shock region. 
The synchrotron spectrum from electrons with the same index $s_e = 1.8$ is consistent with the radio spectral index 
$\alpha \ (\equiv -\Gamma+1) = -0.4 \pm 0.5$ deduced  from the two flux points in Figure~\ref{fig:modeling} \citep{duncan00}. 
Note that the synchrotron spectrum index and electron index are related as $s_e = 2 \Gamma - 1$. 
The flux of the synchrotron component is controlled by the amount of electrons and the magnetic field strength. 
The stronger the magnetic field is, the larger the number of electrons should be, and the inverse Compton flux becomes higher. 
In the hadronic scenario, the magnetic field should be at least $\approx 50~\mu{\rm G}$ so that inverse Compton scattering becomes 
negligible in the GeV-to-TeV domain. 
In the case of $B = 100~\mu{\rm G}$ adopted here, the total kinetic energy of electrons becomes $W_e = 3.9 \times 10^{46}$~erg, and 
the ratio of injected particle numbers at $p = 1~{\rm GeV}~c^{-1}$ is $K_{ep} \equiv A_e/A_p = 1.1 \times 10^{-4}$.

In the case of the leptonic scenario, the electron index is constrained by the \emph{Fermi} LAT spectrum. 
Considering inverse Compton scattering in the Thomson regime, the photon index  and electron 
index are related as $s_e = 2 \Gamma - 1$. 
Therefore, the observed photon index  $\Gamma = 1.85$ leads to a soft electron spectrum with $s_e = 2.7$. 
With this index, however, the synchrotron spectrum cannot fit the radio and X-ray data well.  
We then adopted a harder electron spectrum within the range allowed by the statistical and systematic errors 
in the \emph{Fermi} LAT spectral points. 
The model curves in Figure~\ref{fig:modeling} (b) are the leptonic models when the electron index is $s_e = 2.15$ with 
a high-energy cutoff at $p_{0e} = 25~{\rm TeV}~c^{-1}$. 
The magnetic field strength can be determined as $B = 12~\mu{\rm G}$ so that the synchrotron to inverse Compton flux ratio matches the data. 
Electron bremsstrahlung emission is calculated to be $E^2 dN/dE  \sim 10^{-13}~{\rm erg}~{\rm cm}^{-2}~{\rm s}^{-1}$ at 10~GeV. 

Both models face difficulty. The leptonic model requires a weak magnetic field of order $10~\mu{\rm G}$, 
which may be contradicted by observations. 
The width of the filaments in the shell observed in \emph{Chandra} gives a magnetic field estimate of $\sim 100~\mu{\rm G}$
\citep{bamba05, berezhko09}. 
These results are interpreted as evidence for magnetic field amplification by streaming of accelerated cosmic rays. 
However, It is possible to reconcile a high magnetic field with the leptonic model if GeV gamma rays are radiated not only from the filamentary structures seen by \emph{Chandra} , but also from other regions in the SNR where the magnetic field may be weaker. 

The hadronic model requires an unrealistically large energy of protons if the gas density is as small as 
$\lesssim 0.01~{\rm cm}^{-3}$. 
Thermal X-ray emission works as a probe to estimate the gas density. 
However, no clear detection is reported so far mainly because it is difficult to separate 
emission from RX~J0852.0$-$4622 and the Vela SNR, which overlaps and emits 
strong thermal emission in the soft X-ray band. 
Using \emph{ASCA} data, 
\cite{slane01} placed an upper limit of $n < 3.3 \times 10^{-2}~(d/750~{\rm pc})^{-1/2}~f^{-1/2}~{\rm cm}^{-3}$, where 
$d$ and $f$ are the distance to RX~J0852.0$-$4622 and the filling factor of a sphere taken as the 
emitting volume in the region used for their spectral analysis, respectively. 
In the case of $f \gtrsim 0.4$, which corresponds to rather thick shell width, the upper limit becomes problematic 
since $W_p$ reaches $\gtrsim 10^{51}$~erg, which means that almost all the kinetic energy released by 
the supernova explosion must go to accelerated protons.

The proton index $s_p = 1.8$ for the hadronic scenario is somewhat harder than $s_p = 2.0$ expected 
in the test-particle regime of diffusive shock acceleration with a strong shock. 
Therefore, it is of interest to compare the observed spectrum with models based on diffusive shock acceleration in the nonlinear regime. 
\cite{berezhko09} has calculated the broadband SED of RX~J0852.0$-$4622 
with their time-dependent, nonlinear kinetic theory for cosmic-ray acceleration. 
Their model curve is plotted over the GeV-to-TeV data in Figure~\ref{fig:spec}. 
The \emph{Fermi} LAT spectrum appears harder than the model by \cite{berezhko09}, but the discrepancy 
is not large compared to the systematic errors.

It is of interest to compare the gamma-ray spectrum of RX~J0852.0$-$4622 with that 
of the similar object, RX~J1713.7$-$3946 \citep{fermi1713}. 
The two SNRs have roughly the same age, size, and similar radio, X-ray and TeV gamma-ray spectra. 
The \emph{Fermi} LAT spectrum of RX~J1713.7$-$3946 has a photon index of $\Gamma =1.5 \pm 0.1$, 
which is more preferable for leptonic models than hadronic models \citep{fermi1713}.  
In the case of RX~J0852.0$-$4622, a leptonic origin 
is not favored but cannot be ruled out. 
The difference between the two similar objects could be an important point to be further 
explored to probe gamma-ray production in SNRs. 
Further studies of \emph{Fermi} LAT data in the future, particularly better modeling of the Galactic diffuse emission, 
will reduce the uncertainties and will allow us to probe particle acceleration in the SNR in greater detail.

\acknowledgments
The \emph{Fermi} LAT Collaboration acknowledges support from a number of agencies and institutes for both development and the operation of the LAT as well as scientific data analysis. These include NASA and DOE in the United States, CEA/Irfu and IN2P3/CNRS in France, ASI and INFN in Italy, MEXT, KEK, and JAXA in Japan, and the K.~A.~Wallenberg Foundation, the Swedish Research Council and the National Space Board in Sweden. Additional support from INAF in Italy and CNES in France for science analysis during the operations phase is also gratefully acknowledged.

\end{document}